\documentclass{jetpl}
\twocolumn
\lat
\rtitle{Phase segregation in Na$_x$CoO$_2$\ldots}
\sodtitle{Phase segregation in Na$_x$CoO$_2$ for large Na contents}
\rauthor{T.\,A.\,Platova, I.\,R.\,Mukhamedshin, A.\,V.\,Dooglav, H.\,Alloul}
\sodauthor{Platova, Mukhamedshin, Dooglav, Alloul}
\dates{2010}{}

\abstract{We have investigated a set of sodium cobaltates (Na$_{x}$CoO$_{2}$)
samples with various sodium content (0.67$\leq$x$\leq$0.75) using Nuclear
Quadrupole Resonance (NQR). The four different stable phases and an
intermediate one have been recognized. The NQR spectra of $^{59}$Co allowed us
to clearly differentiate the pure phase samples which could be easily
distinguished from multi-phase samples. Moreover, we have found that keeping
samples at room temperature in contact with humid air leads to destruction of
the phase purity and loss of sodium content. The high sodium content sample
evolves progressively into a mixture of the detected stable phases until it
reaches the $x=2/3$ composition which appears to be the most stable phase in
this part of phase diagram.}

\PACS{81.30.At, 76.60.Gv, 61.66.f, 71.27.+a}

\begin{document}

\title{Phase segregation in Na$_x$CoO$_2$ for large Na contents}
\author{T.\,A.\,Platova$^{1,2}$, I.\,R.\,Mukhamedshin$^{1,2 \,}$\thanks{e-mail: Irek.Mukhamedshin@ksu.ru}, A.\,V.\,Dooglav$^{1}$,\, H.\,Alloul$^{2}$

$^1$Physics Department, Kazan State University, 420008 Kazan, Russia \\
$^2$Laboratoire de Physique des Solides, UMR 8502, Universit\'e Paris-Sud,
91405 Orsay, France}

\maketitle

\textbf{Introduction.} - The family of sodium layered cobaltates
Na$_{x}$CoO$_{2}$ $(0<x\leq 1)$ has a rich phase diagram \cite{Foo}, which
includes most interesting scientific phenomena present in condensed matter
physics, such as superconductivity \cite{Takada}, spin density wave \cite{SDW},
magnetic frustration in a triangular lattice, coexistence of metallic and
magnetic properties, both Curie-Weiss and 2D metal, \emph{etc} \cite{Ivanova}.
Moreover, high ionic mobility and high Seebeck coefficient \cite{HIM,Seebeck}
allow to consider this compound for potential thermoelectric applications
\cite{Bhatt, Moon}.

The concentration $x$ of sodium ions and their order/disorder in the Na plane
play a fundamental role in the physical properties of cobaltates. The Co ions
are in the large crystal field induced by their oxygen octahedral environment,
so the 3d levels are split and the difference in energy between the lower
t$_{2g}$ triplet and upper e$_{g}$ doublet is $\approx $ 2~eV, thus only the
t$_{2g}$ triplet states are filled \cite{Foo}. Therefore the electronic
structure of the Co ions is expected to correspond to low spin configurations
with total electron spin S=0 or S=1/2 with charge states Co$^{3+}$/Co$^{4+}$,
respectively.

In the present work we have studied the cobaltates Na$_{x}$CoO$_{2}$ at large
sodium content $x$ range (0.67$\leq $x$\leq $0.75). This concentration range
bears our attention due to the occurrence of an A-type magnetic ordering at
$x\simeq 0.75$ and its absence at lower sodium contents $x<0.75$
\cite{Foo,Shu}. In Ref.~\cite{EPL2008} the existence of four stable phases in
this Na concentration range has been established. These phases display a
similar nearly ferromagnetic in-plane behavior above 100~K but exhibit
significantly different ground states. The structure of one of these phases has
been proposed recently using NMR/NQR data and confirmed by x-ray Rietveld
analysis \cite{EPL2009, PlatovaPRB}.

In NQR, nuclei with an electric quadrupole moment have their nuclear spin
energies split by the electric field gradient (EFG) created by the electronic
bonds in the local environment. So this technique is very sensitive to the
nature of the bonding around the nucleus.

\textbf{Samples.} - The reproducible preparation of single-phase samples with
precise stoichiometries is not straightforward in cobaltates. The high ionic
mobility of sodium and its chemical activity (for example, Na ions easily react
with molecules present in the ambient atmosphere to form NaOH or sodium
carbonates) make the control of Na content even more difficult.

\begin{figure}[tbp]
\centering
\includegraphics[width=1\linewidth]{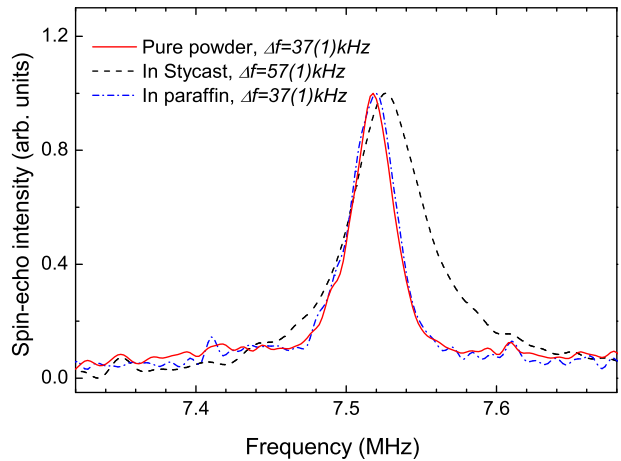}
\caption{Fig.1.(Color online) Cobalt NQR lines of pure powder and powder packed
in Stycast and paraffin: solid (red) line, dotted (black) line and dash-and-dot
(blue) line, respectively.} \label{fig:LineWidth}
\end{figure}

However, the methods for reproducible synthesis of single-phase powder samples
in the 0.67$\leq $x$\leq $0.75 sodium range have been reported in
Ref.~\cite{EPL2008,PlatovaPRB}. To protect the powders from the influence of
water they were packed into protecting materials. Several protecting materials
have been used in our experiments: the epoxy resin (Stycast 1266) and paraffin,
which display distinct advantages and disadvantages. The Stycast perfectly
protects the powder from \textit{water influence}, however we have found that
for samples packed in Stycast, the NQR spectra displayed resonance lines
$\approx $1.9 times broader than pure powder or samples packed in paraffin (see
Fig.~\ref{fig:LineWidth}). Thus, the paraffin packed powder have been used for
most further investigations. To eliminate diffusion processes in the Na layers
the samples were kept in liquid nitrogen.

\textbf{Experimental details.} - The NQR measurements were carried out with a
home-built coherent pulsed NMR/NQR spectrometer. The NQR spectra of $^{59}$Co
were taken \textquotedblleft point by point\textquotedblright with a $\pi
/2-\tau -\pi $ radio frequency pulse sequence by varying the spectrometer
frequency. These sweeps were done with equal frequency steps at 4.2~K. A
Fourier mapping algorithm \cite{Clark, Bussandri} have been used for
constructing the detailed NQR spectra.

\begin{figure*}[tbp]
\centering
\includegraphics[width=0.8\linewidth]{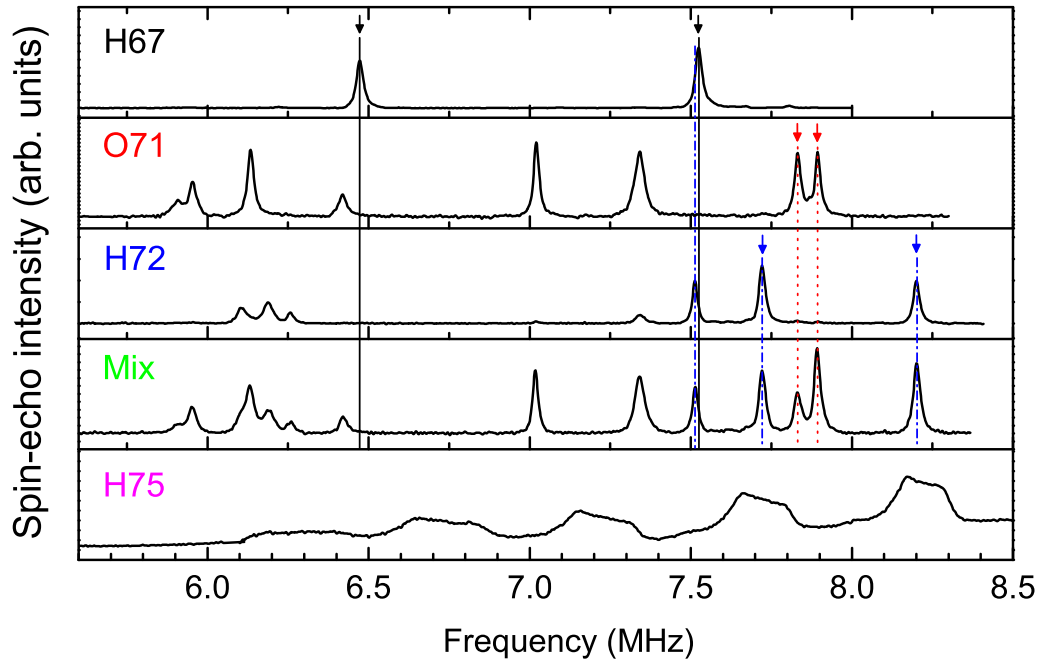}
\caption{Fig.2.(Color online) NQR spectra for the four single phase samples
studied in Ref.~\cite{EPL2008}. The difference in spectra is obvious - each
spectrum has at least one line which does not appear in the other phases as
shown by the coloured arrows and vertical lines (H67-solid (black) line, O71 -
dotted line (red) and H72 -  dash-dotted (blue) lines. The sample denoted as
Mix, with a Na composition in between the O71 and H72 phases, is the mixture of
those phases as its NQR signal is a weighted composition of the NQR signals of
these two phases. H75 which is antiferromagnetically ordered at 4.2~K displays
a broad ZFNMR signal due to the internal magnetic field.}
\label{fig:4phSpe3}
\end{figure*}

\textbf{NQR characterization of the phases}.-We have studied a series of
samples with various sodium contents ($0.67\leq x\leq 0.75$). The NQR spectra
of $^{59}$Co nuclei allowed us to clearly differentiate four stable phases (see
Fig.~\ref{fig:4phSpe3}) marked as H67, O71, H72 and H75. This notation is the
same as used in Ref.~\cite{EPL2008}. The number in the phase label is an
approximate sodium content ($x=0.67$, 0.71, 0.72 and 0.75, respectively) and
the letter is a type of unit cell (H-hexagonal, O-orthorhombic). The part of
the $^{59}$Co NQR spectra which correspond only to the ($\pm $7/2 - $\pm $5/2)
transitions of $^{59}$Co nuclei \cite{EPL2008, PlatovaPRB} are shown in
Fig.~\ref{fig:4phSpe3}. It is clearly seen that every phase has its own unique
$^{59}$Co NQR spectrum. The H67 phase has the simplest spectrum which consists
of two lines at $\approx $6.5~MHz and $\approx $ 7.5~MHz in this frequency
range. In the 5.5$\div $8.5~MHz range the O71 has 8 and H72 has 6 resonance
peaks. The doublet of lines at $\approx $ 7.8~MHz and the two resonance lines
at $\approx $7.75 and $\approx $8.2~MHz are characteristic features of the O71
and H72 phases, respectively. These pairs of lines are labeled in
Fig.~\ref{fig:4phSpe3} by dotted lines and by dash-dotted lines. The spectrum
of H75 phase differs considerably (Fig.~\ref{fig:4phSpe3}) and it will be
discussed below.

Thus, the NQR spectrum is unique and characteristic for each single phase. This
allows easily to distinguish samples which are a mixture of two or more phases.
We show as an example in Fig.~\ref{fig:4phSpe3} the $^{59}$Co NQR spectrum of
the sample (labeled as Mix) with a sodium content intermediate between O71 and
H72  which contains both signals from these two phases. So the NQR is a
sensitive method to distinguish single phase samples from a mixture of phases.
This finding becomes very important as it will allow to characterize better
freshly synthesized samples and clarify the phase diagram of the sodium
cobaltates.

The spectral lines of H67, O71 and H72 phases are rather narrow (linewidths
$\approx $30-50~kHz), which points out that the spread of EFG on the nuclear
site positions is rather small. These phases have a finite number of cobalt
non-equivalent sites indicating the existence of well defined local ordering in
the Co and Na planes. The structural model of the H67 phase
(Na$_{2/3}$CoO$_{2}$ compound) have been proposed recently in
Ref.~\cite{EPL2009,PlatovaPRB}. One unit cell of this models contains four Co
and three Na non equivalent sites which have been detected by NQR
\cite{EPL2009,PlatovaPRB} and by NMR \cite{NaPaper, CoPaper}. From the
comparison of the NQR spectra it is obvious, that the H67 phase has the
simplest structure in the $0.67\leq x\leq 0.75$ sodium concentration range. The
structural organization in the sodium planes is still an open question for the
O71 and H72 phases.

The existence of the antiferromagnetic (AF) order with T$_{N}$=22~K is the
characteristic feature of the $x\simeq 0.75$ compound. Such AF order was
detected by $\mu$SR in our samples \cite{Mendels}, as well as by low field bulk
susceptibility measurements. Neutron scattering study of the same phase
established the A-type AF ordering (ferromagnetic in plane and AF between
planes) \cite{H751,H752}. As our studies have been carried out at 4.2~K, the
H75 phase sample was magnetically ordered at this temperature. Therefore the
observed signal in the H75 phase corresponds to the so-called zero field NMR
(ZFNMR). In this case, the nuclear energy levels are split by the internal
magnetic field. Thus, the observed spectrum consists of seven lines, which
correspond to the typical NMR spectrum for nuclear spin 7/2 (one central line
and 6 satellites), but only five of them are shown on Fig.~\ref{fig:4phSpe3}.
We have failed in detecting the NQR spectrum of the H75 phase above T$_{N}$,
due to significant shortening of transverse relaxation time ($T_{2}$).

\begin{figure}[tbp]
\includegraphics[width=1\linewidth]{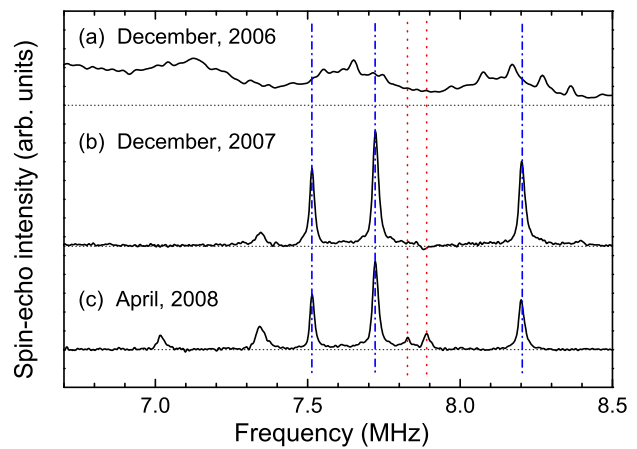}
\caption{Fig.3.(Color online) Evolution of a paraffin packed sample from the
H75 phase (a) to the H72 phase ((b), H72 phase lines are marked with
dash-dotted (blue) lines) and, finally, to the mixture of H72 and O71 phases
((c), O71 phase lines are marked with dotted (red) lines). For details see
text.}
\label{fig:H75evol3}
\end{figure}

During the experiments we have found that the phase content of the samples
packed in paraffin were changed. It should be noted that the paraffin packed
samples between experiments were kept in liquid nitrogen. To perform the
measurements we had to take out samples from liquid nitrogen, warm them to the
room temperature and after that insert them in the probe of our spectrometer.
In Fig.~\ref{fig:H75evol3} the evolution of phase content of one sample after
few tens such thermal cyclings to room temperature is shown. This sample was
initially the H75 single-phase powder (upper spectrum in
Fig.~\ref{fig:H75evol3} - broad ZFNMR lines). However, the powder transformed
progressively into the H72 phase, and then the lines of the O71 phase appeared.
Most probably, during repetitive fast changes of temperature, condensed water
molecules were able to interact with the powder at room temperature.

\begin{figure*}[tbp]
\centering
\includegraphics[width=0.8\linewidth]{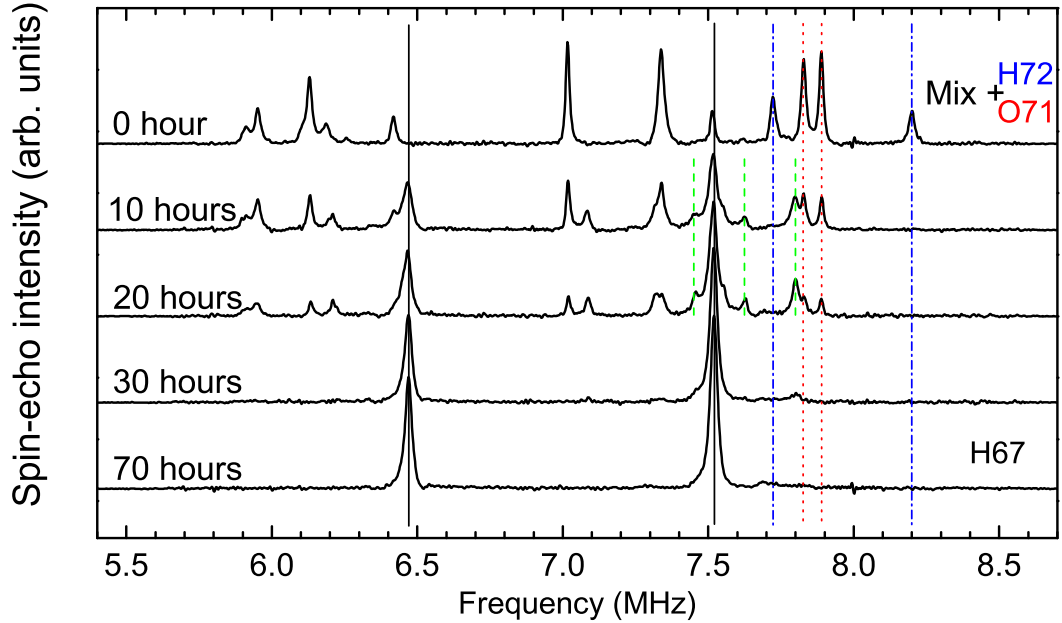}
\caption{Fig.4.(Color online) NQR spectra of $^{59}$Co reflect the evolution of
the phase content of a powder sample versus the time of its exposure to humid
air (see text for details). The sample was initially a mixture of O71 and H72
phases. The phase composition evolves after 10 and 20 hours to O71 + unknown +
H67 phases, while only the H67 phase remains after 30 hours of treatment. This
demonstrates that Na is progressively expelled from the Na$_{x}$CoO$_{2}$
sample.}
\label{fig:Evolution3}
\end{figure*}

\textbf{Phase segregation}.-Therefore the phase composition of the samples
could change due to non-perfect storage conditions at room temperature. To
follow such pocess a special experiment has been performed. A powder sample
which initially was a mixture of H72 and O71 phases has been kept in humid
atmosphere during 3 days at room temperature. The humidity was maintained
$\simeq $75 percent level in the closed half-filled bottle by saturated
solution of NaCl in distilled water. Therefore the powder could easily interact
with the water vapour in air.

We have measured the $^{59}$Co NQR spectrum every 10 hours of exposure powder
sample to humid air, and results are shown in Fig.~\ref{fig:Evolution3}. There
the NQR lines of the known phases are marked by the vertical lines similar to
those in Fig.~\ref{fig:4phSpe3}. The H72 phase had almost disappeared after the
first 10 hours of treatment. At the same time NQR lines in the spectrum
corresponding to the H67 phase appeared, as well as NQR lines of another phase
with intermediate sodium content between 0.71 and 0.67. This unknown and
unstable phase persisted up to the 20th hour of treatment. Also a significant
reduction of the O71 phase and clear increase of H67 phase content was observed
in the 20th hours spectrum (Fig.~\ref{fig:4phSpe3}). After 30 hours of the
sample exposure to humid air the O71 and H72 phases had almost disappeared and
the only remaining phase was the H67 one with slight background of impurity
phases. No significant changes in the $^{59}$Co NQR spectra were detected
during further treatment of the sample in humid air.

However, from our experience it is known that keeping cobaltates powder in air
for a long time leads to the destruction of H67 phase too. We had monitored
such process observing the slow decrease of the H67 phase $^{59}$Co NQR signal
intensity and the appearance of a broad background signal from unknown phases -
see 70 hours spectrum in the Fig.~\ref{fig:4phSpe3}. After such long exposure
of the sample to the humid air white powder, most likely of sodium carbonates,
appeared on the sample surface.

Thus, the powder samples evolve rapidly at room temperature in contact with
humid air. The evolution process indicates a reduction of sodium content
associated with a loss of Na ions. This statement is in a very good agreement
with former work by Shu \textit{et al.} \cite{Shu} done at even higher Na
content. These authors have investigated a single crystal Na$_{0.88}$CoO$_{2}$
and revealed a loss of sodium ions from the surface which were fixed in a white
powder appearing on the crystal surface. Apparently, the water and/or CO$_{2}$
molecules react with Na ions from surface with formation of sodium hydroxide
(NaOH) and sodium carbonates. So to avoid as much as possible the change of
phase composition of the powder, we recommend to isolate cobaltates powders
from interaction with humid air and to try to block the diffusion of Na ions.
In some sense, the best way to store the cobaltates powders is to keep them at
low temperature in hermetic ampoules with a small amount of helium gas as a
heat exchanger. This arrangement isolates the powder from interaction with
atmosphere and allows to perform experiments at low temperature.

\textbf{Conclusion}.-Low temperature NQR is a very powerful method to
investigate the phase content of sodium cobaltates. The four stable phases and
an intermediate unstable one have been detected in the $0.67\leq x\leq 0.75\
$sodium range. The spectral lines of phases without magnetic ordering are
rather narrow and non-overlapping, and the spectra could be described fully by
finite set of Co non-equivalent sites. All these factors point out the
existence of well defined orders in the Co and Na planes. These results are in
good agreement with previous investigations \cite{EPL2008}. So far, only the
structure of the H67 phase has been determined \cite{PlatovaPRB}, confirmed by
x-rays \cite{EPL2009} and agrees with LDA computations \cite{Meng}. The present
NQR results in combination with diffraction data should help to clarify the
structures of the other phases. This should shed some light on the origin of
the Curie-Weiss susceptibility behavior of studied phases and magnetic ordering
of $x\simeq 0.75$ phase.

We thank G.~Collin and N.~Blanchard for synthesizing the samples and for
fruitful discussions. This study was partly supported by the RFBR under project
10-02-01005.

\end{document}